\documentclass[]{aa}

\usepackage{graphicx}
\usepackage{txfonts}
\usepackage[]{hyperref}

\usepackage{natbib}
\usepackage{multirow}
\usepackage{multicol}
\usepackage{textcomp}
\usepackage{xcolor}
\usepackage{verbatim}
\usepackage[labelfont=bf]{caption}
\begin{document}

\title{ISPY - the NaCo Imaging Survey for Planets around Young stars}
\subtitle{A young companion candidate embedded in the R\,CrA cloud\thanks{Based on observations collected at the Paranal Observatory, ESO (Chile). Program ID: 199.C-0065(A) and 1101.C-0092(A)}/\thanks{After this paper was first submitted, another study of R CrA corroborating the reality of its companion, and reaching conclusions largely in agreement with ours, was posted on arXiv \citep{Mesa2019}}}

\author{G. Cugno\inst{\ref{ethz}}
\and S.~P. Quanz\inst{\ref{ethz},\ref{nccr}}
\and R. Launhardt\inst{\ref{MPIAh}}
\and A. Musso Barcucci\inst{\ref{MPIAh}}
\and S.~S. Brems\inst{\ref{ZAH}} 
\and A. Cheetham\inst{\ref{geneva}}
\and N. Godoy\inst{\ref{Chile}, \ref{Chile2}}
\and G.~M. Kennedy\inst{\ref{Warwick1}, \ref{Warwick2}}
\and T. Henning\inst{\ref{MPIAh}}
\and A. M\"uller\inst{\ref{MPIAh}}
\and J. Olofsson\inst{\ref{Chile}, \ref{Chile2}}
\and F. Pepe\inst{\ref{geneva}}
\and A. Quirrenbach\inst{\ref{ZAH}}
\and S. Reffert\inst{\ref{ZAH}}
\and E.~L. Rickman\inst{\ref{geneva}}
\and D. S\'egransan\inst{\ref{geneva}}
}

\institute{ETH Zurich, Institute for Particle Physics and Astrophysics, Wolfgang-Pauli-Strasse 27, CH-8093 Zurich, Switzerland\label{ethz}
\and National Center of Competence in Research "PlanetS" (\url{http://nccr-planets.ch})\label{nccr} 
\and Institute for Astronomy (MPIA), K\"onigstuhl 17, 69117 Heidelberg, Germany \label{MPIAh}
\and Landessternwarte, Zentrum f\"ur Astronomie der Universit\"at Heidelberg, K\"onigstuhl 12, 69117 Heidelberg, Germany \label{ZAH}
\and Observatoire Astronomique de l’Universit\'e de Gen\`eve, 51 Ch. des Maillettes, 1290 Versoix, Switzerland\label{geneva}
\and Instituto de F\'isica y Astronom\'ia, Facultad de Ciencas, Universidad de Valpara\'iso, Av. Gran Breta\~na 1111, Playa Ancha, Valpara\'iso, Chile\label{Chile}
\and N\'ucleo Milenio Formaci\'on Planetaria - NPF, Universidad de Valpara\'iso, Av. Gran Breta\~na 1111, Valpara\'iso, Chile\label{Chile2}
\and Department of Physics, University of Warwick, Gibbet Hill Road, Coventry, CV4 7AL, UK\label{Warwick1}
\and Centre for Exoplanets and Habitability, University of Warwick, Gibbet Hill Road, Coventry, CV4 7AL, UK\label{Warwick2}
\\\\\email{gabriele.cugno@phys.ethz.ch}
}

%

\date{Received --- ; accepted --- }

\abstract{}{}{}{}{}

\abstract
{Within the NaCo-ISPY exoplanet imaging program, we aim at detecting and characterizing the population of low-mass companions at wide separations ($\gtrsim$10AU), focusing in particular on young stars either hosting a known protoplanetary disk or a debris disk.}
{R\,CrA is one of the youngest (1-3 Myr) and most promising objects in our  sample because of two previous studies that suggested the presence of a close companion. Our aim is to directly image and characterize the companion for the first time. }
{We observed R\,CrA twice with the NaCo instrument at VLT in the $L'$ filter with a one year time baseline in between. The high-contrast imaging data were reduced and analyzed, and in both datasets the companion candidate was detected. We used artificial negative signals to determine the position and brightness of the companion and the related uncertainties. }
{The companion is detected at a separation of $196.8\pm4.5$/$196.6\pm5.9$ mas ($18.7\pm1.3$/$18.7\pm1.4$ AU) and position angle of $134.7\pm0.5^\circ/133.7\pm0.7^\circ$ in the first/second epoch observation. We measure a contrast of $7.29\pm0.18$/$6.70\pm0.15$ mag with respect to the primary. Stellar proper motion study rejects the hypothesis of the signal being a background object. The companion candidate orbits in the clockwise direction and, if on a face-on circular orbit, its period is $\sim43-47$ yr. This value disagrees with the estimated orbital motion and therefore a face-on circular orbit may be excluded.  Depending on the assumed age, extinction and brightness of the primary, the stellar companion has a mass between $0.10\pm0.02\, M_\odot$ and $1.03^{+0.20}_{-0.18}\,M_\odot$ range, if no contribution from circumsecondary material is taken into account.}
{As already hypothesized by previous studies, we have directly detected a low-mass stellar companion orbiting the young Herbig Ae/Be star R\,CrA. Depending on the age assumptions, the companion is among the youngest forming companions imaged to date, and its presence needs to be taken into account when analyzing the complex circumstellar environment of R\,CrA.}


\titlerunning{A young companion candidate embedded in the R\,CrA cloud}
\maketitle


\section{Introduction}
\label{section1}
In recent years, high-contrast high spatial resolution direct imaging has revealed a few planetary mass objects \citep[$\beta$ Pic\,b, HR\,8799\,bcde, HD\,95086\,b, 51\,Eri\,b, HIP\,65426\,b, PDS\,70\,b; ][]{Lagrange2008, Marois2008, Marois2010, Rameau2013, Macintosh2015, Chauvin2017, Keppler2018}. Several surveys \citep[such as NICI, SPHERE-SHINE, IDPS, SEEDS, LEECH and GPI; ][]{Liu2010, Chauvin2017_shine, Vigan2012, Tamura2014, Skemer2014, Macintosh2014} exploit the capabilities of existing instrumentation to put stronger constraints on the occurrence rate of giant planets ($5-13\,M_J$) at large orbital distance ($30-300$ AU), which was estimated to be $0.6^{+0.7}_{-0.5}\%$ by \cite{Bowler2016}. Statistically characterizing the giant planet population in wide orbits (>10 AU) is also one of the primary goals of the NaCo Imaging Survey for Planets around Young stars (NaCo-ISPY, Launhardt et al., in prep). The survey targets nearby young stars which are surrounded by either a protoplanetary or a debris disk. Unlike most other ongoing imaging surveys that focus on shorter wavelengths, ISPY utilizes the $L'$ filter ($\lambda\sim3.8\,\mu$m) of the NaCo instrument at the Very Large Telescope (VLT) to obtain high-contrast imaging data. At this wavelength, there are four main advantages compared to the near-infrared regime: (1) the contrast requirements are more favorable for a detection, (2) dust extinction from circumstellar material has minimal effects on the observed flux, (3) scattering from circumstellar disk material, which increases the noise and the probability of False Positives, is less important and (4) contamination from background objects is less severe because most background stars are not red enough to be easily detected at $L'$ band. Altogether, possibly embedded objects with lower intrinsic luminosity might be detected. The major limiting factor of collecting data in the 3-5 $\mu$m range is related to the lower spatial resolution that can be achieved.\\
Given its age and proximity (see Table \ref{tab:object}), R\,CrA is an ideal target for the ISPY search for companions at small physical separations. 
R\,CrA is located in the Coronae Australis  molecular cloud, a star-forming region containing several low-to-intermediate mass objects \citep[e.g., ][]{FernandezComeron2001, Haas2008}. Its `Coronet' \citep{TaylorStorey1984, WIlking1997} is a deeply embedded cluster of young stars that was extensively studied in the past years across all wavelengths, from radio \citep[e.g., ][]{Brown1987, Feigelson1998, Forbrich2006} to X-rays \citep[e.g., ][]{Koyama1996, Neuhauser2000}, and whose luminosity is dominated by R\,CrA \citep[$L_{\rm bol}=70\,L_\odot$,][ but updated with distance from Table \ref{tab:object}]{Lorenzetti1999}. Because of the complexity of its circumstellar environment and its known variability, the properties of R\,CrA are not well constrained (see Table \ref{tab:object}).
\begin{table}[t!]
\caption{\label{tab:object} Stellar parameters of R\,CrA.}
\centering
\begin{tabular}{llc}
\hline\hline\noalign{\smallskip}
Parameter		& 	Values	& Reference \\\hline
\noalign{\smallskip}
RA & 19$^h$01$^m$53.68$^s$ & (1) \\
\noalign{\smallskip}
DEC & -36$^\circ$57$'$08.14$''$ & (1) \\
\noalign{\smallskip}
Sp. type 	& F5, A5, B8 & (2), (3), (4)		\\
\noalign{\smallskip}
Dist. [pc]  & $94.9\pm6.3$& (1)\\
\noalign{\smallskip}
$\mu_\alpha\times\cos(\delta)$ [mas/yr] & $1.58\pm1.20$& (1)\\
\noalign{\smallskip}
$\mu_\delta$ [mas/yr]  & $-30.83\pm1.19$& (1)\\
\noalign{\smallskip}
$K$ [mag]  & $2.858\pm0.262$& (5)\\
\noalign{\smallskip}
$L$ [mag]  & $1.78\pm0.05$& (6)\\
\noalign{\smallskip}
WISE $W3$ [mag]  & $-1.239\pm0.390$& (7)\\
\noalign{\smallskip}
Age [Myr] & 0.3-3, 1 & (8), (9), (10)\\
\noalign{\smallskip}
$A_V$ [mag]   & $6.0-8.0$, 1.9 & (9), (11) \\
\noalign{\smallskip}
Mass [$M_\odot$]   & 3.0, 3.5 & (10), (11) \\
\noalign{\smallskip}
\hline\hline
\end{tabular}
\tablebib{(1) \cite{Gaia2018}, (2) \cite{Hillenbrand1992}, (3) \cite{Chen1997}, (4) \cite{Hamaguchi2005}, (5) \cite{Cutri2003}, (6) \cite{Glass1975}, (7) \cite{Cutri2013}, (8) \cite{Meyer2009}, (9) \cite{Forbrich2006}, (10) \cite{Bibo1992}, (11) \cite{Lorenzetti1999}.}
\end{table}
As an example, the extinction in the R\,CrA cloud is highly uncertain, and X-ray spectra from \cite{Forbrich2006} suggested $A_V=6.0-8.0$ mag, but other studies derived lower values \citep[e.g., ][$A_V=1.9$ mag]{Lorenzetti1999}. Since it was not possible to further constrain this parameter from the available data, we adopted a conservative approach considering $A_V$ in the $1.9-8.0$ mag range.
Concerning possible companions, \cite{Takami2003} observed R\,CrA with the Royal Greenwich Observatory (RGO) spectrograph on the Anglo-Australian Telescope (AAT) and the data showed a displacement around the H$\alpha$ line emission that was best explained with the presence of a companion and an outflow from the primary. Combining their results with previous observations presented in \cite{Bailey1998}, they gave a rough estimate for the position angle (PA) of a possible companion in 1996 and 1999 ($\sim220^\circ$ and $\sim200^\circ$). Later, \cite{Forbrich2006} hypothesized that a companion was orbiting R\,CrA based on  \textit{Chandra} and XMM-\textit{Newton} data. They detected  variable X-ray emission but because no X-ray flux is expected from intermediate mass young stellar objects, they suggested that the emission comes from a low-mass companion. \cite{Kraus2009} analyzed VLTI/AMBER $H-$ and $K-$band spectro-interferometric observations of R\,CrA and found no clear evidence for a companion at separations between $0\farcs02$ and $0\farcs2$, despite strong asymmetries in the source's brightness distribution that could be caused by a second object. They achieved contrast limits of 4.0 mag, so a fainter (i.e., lower mass or more deeply embedded) object would have been missed in their observations. Furthermore, they measured the inclination of the disk surrounding R\,CrA to be $\sim35^\circ$, \citep[confirming previous results from, e.g.,][]{Clark2000} and its position angle to be $\sim180^\circ-190^\circ$.

In this paper, we present the results of new high-contrast imaging $L'$-band observations of R\,CrA that find direct evidence for a stellar companion. In Section \ref{section2} we summarize the observations and explain the data reduction. In Section \ref{section3} we present our results and discuss them in Section \ref{section4}. We conclude in Section \ref{section5}.\\


\section{Observations and data reduction}
\label{section2}
\subsection{VLT/NaCo observations}
\begin{figure*}[t!]
\centering
\includegraphics[width=\hsize]{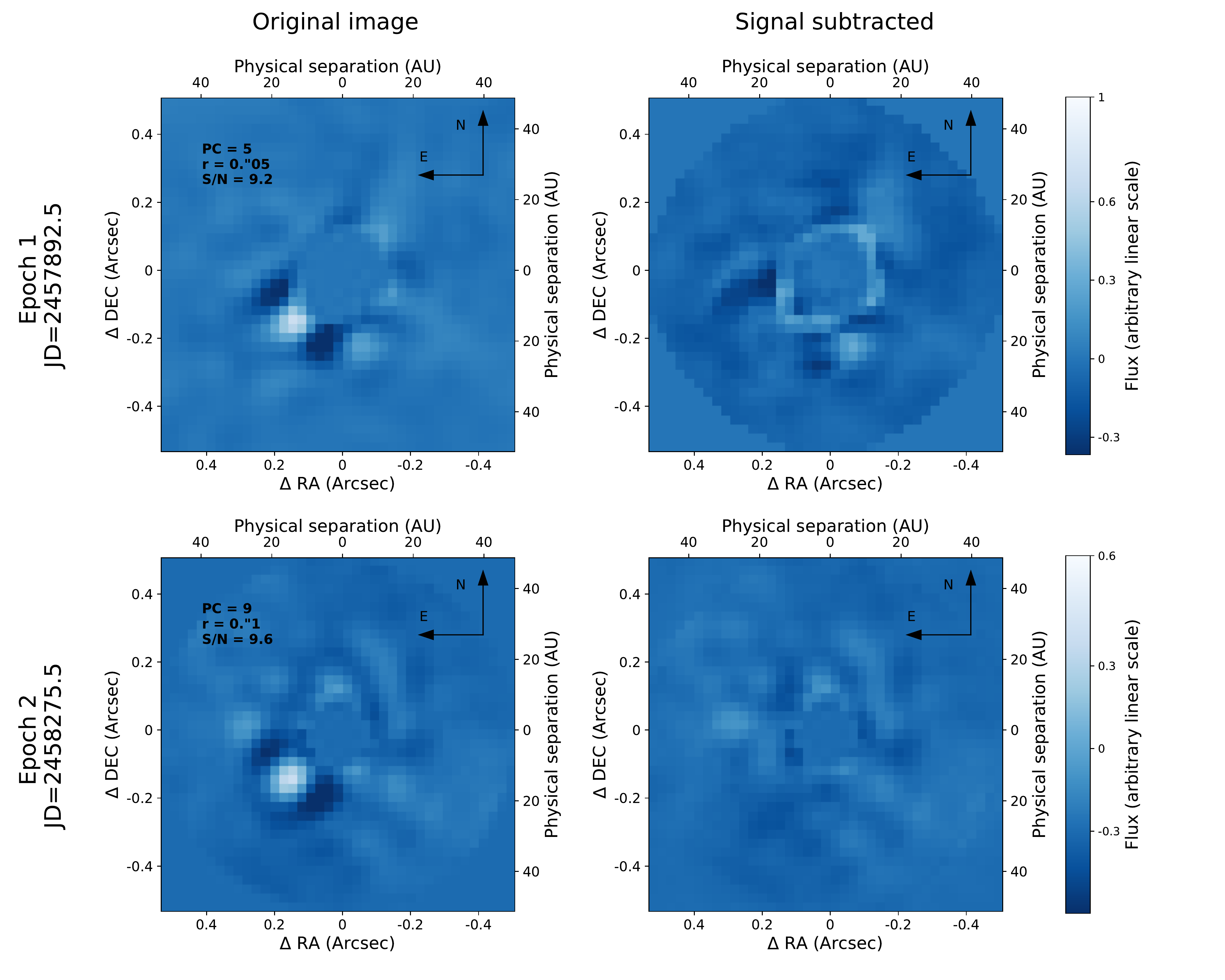}
\caption{First and second epoch detections (top and bottom panels respectively) with the highest S/N (left) and the corresponding signal-subtracted images (right). The numbers of subtracted PCs from the central PSF, the applied mask radius and the SNR are reported on the top left corner of the original images. The same parameters were used for the right panels, where prior to PSF subtraction the Hessian Matrix approach was used to subtract the companion signal.}
\label{fig:RCrA_images}
\end{figure*}

\begin{table}[t!]
\caption{\label{tab:observations} Summary of observations.}
\centering
\begin{tabular}{lll}
\hline\hline\noalign{\smallskip}
Parameter  		& Epoch 1				& Epoch 2 \\\hline
\noalign{\smallskip}
UT date [yyyy/mm/dd]   & 2017/05/19    & 2018/06/06  \\
\noalign{\smallskip}
Prog. ID   & 199.C-0065(A)    & 1101.C-0092(A)  \\
\noalign{\smallskip}
DIT\tablefootmark{a} [s]   & 0.35    & 0.1082  \\
\noalign{\smallskip}
\# of DITs   & 12100    & 18401  \\
\noalign{\smallskip}
Par. angle start/end [$^\circ$]  & 42.46/78.44    & -40.25/80.91  \\
\noalign{\smallskip}
Airmass min/max  & 1.04-1.19  & 1.04-1.22  \\
\noalign{\smallskip}
Mean seeing\tablefootmark{b} [as|   & 0.65    & 0.63  \\
\noalign{\smallskip}
True North\tablefootmark{c} [$^\circ$] & $0.39\pm0.21$ & $0.43\pm0.18$\\
\noalign{\smallskip}
Plate Scale [mas/pix] & $27.21\pm0.10$ & $27.21\pm0.08$\\
\noalign{\smallskip}
\hline
\noalign{\smallskip}
\# unsaturated PSF & 2000 & 2400\\
\noalign{\smallskip}
DIT unsaturated PSF [s] & 0.005 & 0.004256\\
\noalign{\smallskip}
\hline\hline
\end{tabular}
\tablefoot{\tablefoottext{a}{DIT = Detector integration time, i.e., exposure time per image frame.}\tablefoottext{b}{Mean DIMM seeing measured during the observations.} \tablefoottext{c}{Additional rotation in clockwise direction that has to be applied in order to have north pointing to the top of the images.}}
\end{table}
We observed R\,CrA on 2017-05-19 and 2018-06-06, taking advantage of the AGPM vortex vector coronagraph \citep{mawet2013} that enables high contrast imaging very close to the diffraction limit ($\sim0\farcs095$). For the observations, the L27 camera (pixel scale $\sim 27.2$ mas pix$^{-1}$) was used with the $L'$ filter ($\lambda_c=3.8\,\mu$m, $\Delta \lambda =0.62\,\mu $m). The camera derotator was switched off to enable Angular Differential Imaging \citep[ADI,][]{marois2006} and the `cube mode' allowed us to save all individual exposures. Both observing runs consisted of series of 13 coronagraphic cubes of images of R\,CrA (7.6 and 2.3 minutes for the two epochs) interposed with one cube of sky images to sample the background and subtract it later. In total, for the first and the second observations we obtained 70.6/33.2 minutes of on-source integration time and 36.0/121.2 degrees of field rotation (see Table \ref{tab:observations}). In addition, unsaturated photometric images were taken at the beginning and at the end of the observations. These were used later to calibrate, simulate and insert artificial PSFs in the data. Table \ref{tab:observations} summarizes the observations together with the atmospheric conditions. 

\subsection{Data reduction}

Both datasets were reduced with \texttt{PynPoint} \citep{amaraquanz2012, Stolker2018}, an end-to-end modular python pipeline for analyzing high-contrast imaging data. In particular, the PSF subtraction is based on Principal Component Analysis \citep[PCA, ][]{amaraquanz2012, Soummer2012}.\\
After assigning to each frame its parallactic angle by linearly interpolating between the start and end value of each cube, the first 10 frames and the last frame of each cube were removed. The first frames suffer from a systematically higher background that drops exponentially \citep[e.g.,][]{Stolker2018}. The last frame represents the mean of the cube. The remaining frames were dark-subtracted, and the stripes present in the lower left quadrant due to detector issues were corrected by substituting each pixel value with the mean of the two neighboring values. For the second epoch, where two neighboring bad working columns were present on the detector, the average of the closest functioning right and left pixels were used\footnote{We note, however, that the stripe correction has no influence on the results presented later, since the studied region was located in a different quadrant.}. Other bad pixels were corrected by 4-sigma clipping \citep{Stolker2018}, then using spline interpolation all images were aligned applying cross-correlation and the entire stack was shifted to the exact image center after fitting a 2D Gaussian to the mean of the frames. Next, each background cube was averaged  to a single image. From each science frame, the average between the previous and the subsequent background image was subtracted. Finally, we binned, i.e., mean-combined, sets of 2, 5 and 10 frames. In this way, three different datasets were created, and a frame selection was applied to each of the three stacks rejecting images where the stellar flux measured in a circular aperture (r =  $1\farcs0$) deviated by more than 3 standard deviations from the median values of the stack. 

For the PSF subtraction, \texttt{PynPoint} applies full frame PCA to reconstruct and then remove the stellar PSF before de-rotating and combining all the images \citep{amaraquanz2012, Stolker2018}. In order to search for faint high-contrast signals at small separations, several mask radii are applied to cover the central part of the stellar PSF and the number of principal components (PC) used to model and subtract the PSF was varied. \\
\begin{figure*}[t!]
\centering
\includegraphics[width=\hsize]{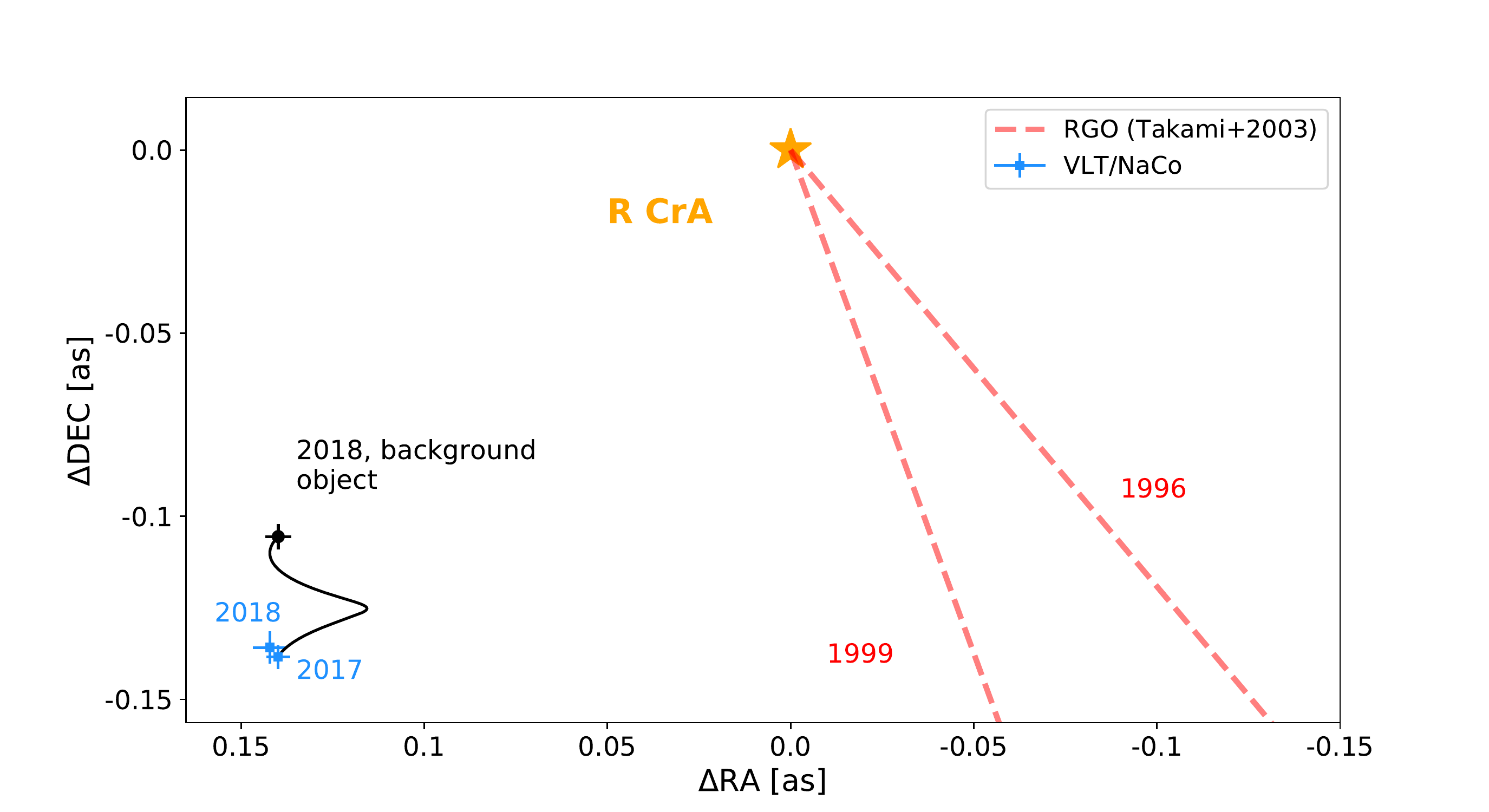}
\caption{Astrometry chart of R\,CrA and its companion candidate. The orange asterisk represents the position of R\,CrA, the blue squares represent the detected relative positions presented in this work, while the black circle shows the expected position of a stationary background object at the second epoch, given the parallax and proper motion of R\,CrA as reported in Table \ref{tab:object}. The red dashed lines represent the position angles at which the companion was predicted by \cite{Takami2003} from 1996 and 1999 data.}
\label{fig:proper_motion}
\end{figure*}
The unsaturated PSF frames underwent a simpler calibration. After dark subtraction and bad pixel cleaning, the background was subtracted using the dithering technique, in which the star is imaged in three different quadrants of the detector\footnote{We avoided using the quadrant with bad columns}. Whenever the star was not present in a quadrant, these cubes served to sample the background. Because the dither position changed every 400 frames, the images were subtracted from each other with an offset of 400 frames, meaning that the $n^{th}$ image of the set was subtracted from the $(n+400)^{th}$ frame and cut ($0\farcs5$ in size) around the positive PSF signal. Then, all frames were aligned, centered and averaged to obtain a unique PSF template for each epoch.\\
We looked for differences in the unsaturated images within individual epochs and between them. Since the final PSF template is unique for each dataset, we have to consider variations of the unsaturated images due to changes in weather conditions and possibly in intrinsic stellar flux. This component of the photometric uncertainty was estimated as the standard deviation of the stellar flux measured in each exposure in a circular aperture of $0\farcs19$ (2 $\lambda/D$) and it amounts to 0.03 mag for epoch 1 and 0.09 mag for epoch 2. The larger variation in epoch 2 results from a systematic decrease in stellar flux by $\sim13\%$ between the beginning and the end of the observing sequence. Furthermore, during the first epoch, the central star R\,CrA appeared on average 0.22 mag brighter than during the second observation\footnote{This value is based
on the measured count rate, even though in both nights the conditions
were not photometric.}. This difference will be analyzed in the context of the companion brightness in Section \ref{section4}.

\section{Analysis and results}
\label{section3}
\subsection{The detection of a young companion candidate}
A strong companion signal is consistently detected in both datasets (see Figure \ref{fig:RCrA_images}) and, given the baseline of one year, the displacement is in agreement with orbital motion as we will discuss in Section \ref{section4}. The companion candidate is detected for a wide range in the number of subtracted PCs (1-20 for Epoch 1 and 4-30 for Epoch 2) applying inner masks of  $0\farcs05$ and $0\farcs1$ in radius. To quantify the confidence of the detection, we calculated the False Positive Fraction (FPF) measuring the signal flux in an aperture of diameter $\lambda/D$ ($\approx0\farcs095$) and sampling the residual noise in non-overlapping apertures of the same size at the same separation, but at different azimuthal angles \citep{mawet2014}. The central position of the signal aperture was determined by fitting a 2D Gaussian to the signal. Using equations 9 and 10  of \cite{mawet2014} we determined the Signal-to-Noise-Ratio (S/N) and the FPF. This approach considers corrections for small sample statistics and utilizes the Student t-distribution to estimate FPFs. We found a best S/N of 9.2 for the first epoch (no frame binning, PC=5, $r_{mask}=0\farcs05$ as) and 9.6 for the second one (bin=10, PC=9, $r_{mask}=0\farcs1$), corresponding to FPFs of $1.7\times10^{-6}$ and $1.17\times10^{-6}$, respectively.

\subsection{Astrometry and photometry}
To determine the astrometry and the photometry of the companion, two different methods were used, namely the minimization of the determinants of the Hessian Matrix (HM) at the position of the signal \citep{Cugno2018} and an MCMC minimization algorithm \citep{Stolker2018}. Both methods are based on the artificial planet technique, in which signals from unsaturated PSFs are subtracted from the data with different combinations of position and contrast to minimize the residuals. Specifically, the HM technique tries to minimize the curvature of the image in the signal region, the MCMC minimizes the flux residuals as performed in \cite{Wertz2017}. In both cases a circular region with $r=0\farcs1425$ (1.5 $\lambda/D$) around the companion was considered.\\
We used the HM characterization algorithm on all combinations of PCs and mask radius that yielded an FPF $<3.2\times10^{-5}$ (corresponding to a 4$\sigma$ confidence), in the bin=5 and bin=10 datasets (22 setups for epoch 1 and 56 for epoch 2). To consider systematic effects related to the PSF subtraction parameters, we averaged the best fit solutions to obtain astrometry and photometry of the companion. The uncertainties are represented by the standard deviation of the set of best solutions. \\The MCMC characterization algorithm was performed for stacks with 10 binned images using the parameter combination that delivered the lowest FPF, namely PC=3, mask=$0\farcs05$ for epoch 1, and PC=9, mask=$0\farcs1$ for epoch 2. The parameter space was explored by 200 walkers, each performing a chain of 500 steps.\\
Astrometric calibration was performed using an astrometric reference field in 47\,Tuc taken on 2017-05-19 and on 2018-06-07, and comparing it to the catalog of \cite{McLaughlin2006}. The resulting True North correction and platescale are summarized in Table \ref{tab:observations}.
The measured contrasts were corrected for the separation-dependent throughput of the coronagraph \citep{mawet2013}, which was 0.91 for both epochs. 
Table \ref{tab:astrometry} summarizes the measurements of the companion for both methods and epochs after those corrections, still without considering photometric variability. The formal errors obtained through the MCMC algorithm do not include any systematic uncertainty and therefore are much smaller than those obtained for HM.
For epoch 1, the astrometric results obtained through both methods agree within 1-$\sigma$ errorbars, while the measurements obtained for the data of epoch 2 show a larger scatter. For both cases, the MCMC approach measured a lower contrast than the HM. To investigate the presence of systematics, we inserted signals at $0\farcs2$ at 3 different position angles randomly chosen (but separated by at least 50$^\circ$ from the companion) with 7.3 mag contrast. The analysis shows that neither of the two approaches systematically delivers more accurate results; sometimes the HM algorithm gives values closer to the known ground-truth, sometimes the MCMC sampling. This is most probably due to the presence of residual PSF speckle noise in the image that affect the measurements. Because the HM approach includes at least in part systematic uncertainties, we decided to use these results for the remainder of our analysis. 

\begin{table}[t!]
\caption{\label{tab:astrometry} Measured astrometry and photometry of R\,CrA\,B.}
\centering
\begin{tabular}{llll}
\hline\hline\noalign{\smallskip}
Epoch 1 		& 	Sep [mas]	& PA [$^\circ$] & Contrast [mag] \\\hline
\noalign{\smallskip}
Hessian matrix  & $196.8\pm4.5$    & $134.7\pm0.5$ & $7.29\pm0.18$ \\
\noalign{\smallskip}
MCMC   &   $192.9\pm1.0$  & $135.0\pm0.2$ &  $6.93^{+0.01}_{-0.02} $ \smallskip\\\hline\hline
\noalign{\smallskip}
Epoch 2 & & & \\\hline
\noalign{\smallskip}
Hessian matrix   & $196.6\pm 5.9$   & $133.7\pm0.7$  & $6.70\pm0.15$\\
\noalign{\smallskip}
MCMC   & $186.7\pm 1.1$    & $132.0\pm 0.2$  & $6.48\pm0.01$\\
\noalign{\smallskip}
\hline\hline
\end{tabular}
\end{table}

\section{Discussion}
\label{section4}
\begin{figure}[t!]
\centering
\includegraphics[width=\hsize]{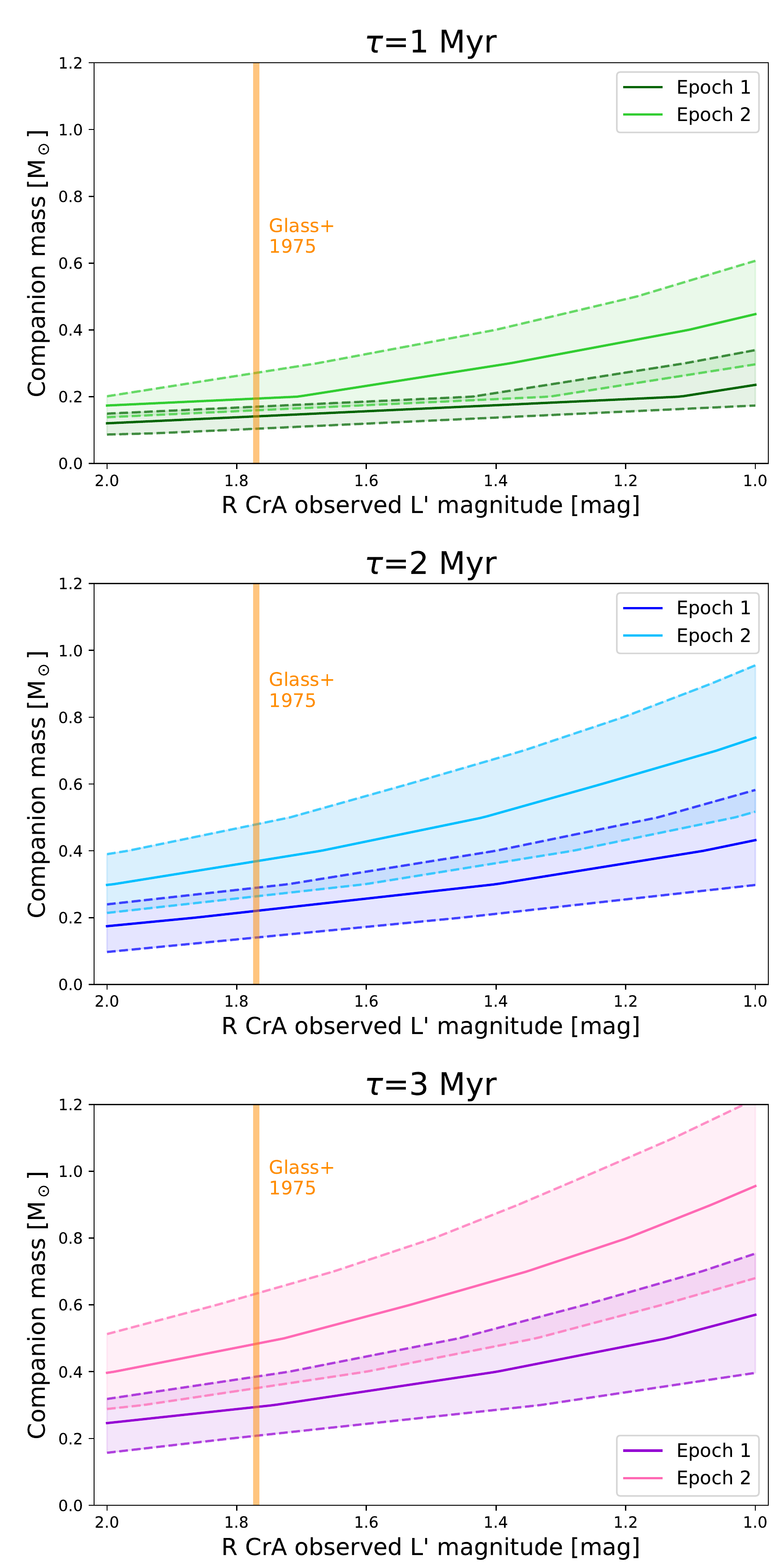}
\caption{Companion mass as a function of the primary $L'$ magnitude for $\tau=1,2,3$ Myr (top to bottom). The shaded regions represent the 1$\sigma$ uncertainty range coming from contrast, photometric uncertainty of the PSF template, distance and extinction. The orange solid line indicates the value provided by \citet{Glass1975}.}
\label{fig:mass}
\end{figure}
\subsection{Exclusion of a background source}
In Figure \ref{fig:proper_motion} we show in blue squares the relative astrometric position of R\,CrA B, keeping the primary at a fixed position. The black line and circle represent the path and the position we would  expect for a stationary background object according to the \textit{Gaia} proper motion of R\,CrA, which is inconsistent with the observations. Error bars represent uncertainties for the astrometric results obtained with the Hessian Matrix approach and include also the uncertainties in \textit{Gaia} parallax and proper motion for the black circle. However, both the Gaia DR2 astrometric excess noise of 3.3\,mas, as well as the large Unit Weight Error of \,$\approx 31$ \citep[UWE, the currently most useful goodness-of-fit indicator for the Gaia DR2 five-parameter astrometric solutions;][]{Lindegren2018}, indicate that the single star astrometric solution does not fit the observations well. These inconsistencies could be due to systematic errors not taken into account by the formal uncertainties, but at the given level it is more likely that the companion to R\,CrA is responsible for the larger than expected residuals to the single star solution.\\
Combining predictions from \cite{Takami2003} with the new information from NaCo observations, it seems that the secondary source moves west to east and, if bound, it orbits R\,CrA in a clockwise direction (current projected separation $\sim18.7$ AU). 
Assuming a face-on circular orbit, its period would be $\sim43-47$ yr (central star mass $\sim3.0-3.5\,M_\odot$). This value implies an orbital velocity of $\sim 7.7-8.3$ deg/yr. Considering the PAs estimated between 1996 and 2018, the orbital velocity was calculated to be $\sim4.0$ deg/yr, while the one between the ISPY detections is $1\pm0.9$ deg/yr. Therefore, based on the 22 years long baseline, the inconsistency allows us to marginally exclude a face-on orbit. Given the lack of uncertainties, confidence levels and orbital separations in the predictions from \cite{Takami2003}, it is not possible to draw more significant conclusions and more precise constraints from the available information.

\subsection{Mass of the companion candidate}
\label{sec:Mass}
 To estimate the $L'$ brightness of R\,CrA, we considered the 2MASS \citep{Cutri2003} $K$-band photometry and the one from the AllWISE Catalog \citep{Cutri2013} in $W3$ ($\lambda\sim11.6\,\mu$m) as reported in Table \ref{tab:object}. We estimated fluxes in the two bands using filter zero points\footnote{\url{http://svo2.cab.inta-csic.es/theory/fps3/}}, and interpolated between the two values. The resulting flux was translated into magnitude and we obtained a brightness of $1.5\pm0.5$ mag at $3.8\,\mu$m. This method avoids using the highly saturated and therefore rather unreliable measurements taken with the WISE \citep{Wright2010} $W1$ and $W2$ bands. This estimate agrees with the brightness measured by \cite{Glass1975}, which is $1.78\pm0.05$ mag, where the error is an upper limit for the uncertainty. To be conservative, we considered $L'$  values for the R\,CrA system to be in the $2.0-1.0$ mag range. It is likely that the measured infrared emission is affected by contribution from the circumstellar disk and surrounding material. Nevertheless, the contrast has been measured with respect to the observed flux, and the intrinsic stellar brightness is not relevant for the characterization of the companion.
After combining the primary brightness with the measured contrasts, we corrected for extinction ($A_{V}=1.9-8.0$ mag, Tab.~\ref{tab:object}) using the extinction curve from \cite{Mathis1990} for $R_V=3.1$. This resulted in $A_{L'}=0.09-0.36$ mag. Afterwards, absolute magnitudes for the companion candidate were computed and compared with the interpolated isochrones from the BT-Settl models \citep{Baraffe2003} for ages of $\tau=1, 2$ and 3 Myr. 
As shown in Figure \ref{fig:mass}, depending on the brightness, extinction and age of the primary at the time of the observations, the companion could have masses ranging from $0.10\pm0.02\,M_\odot$ to $1.03^{+0.20}_{-0.18}\,M_\odot$.
Considering the particular case in which the primary's brightness was equal to $1.78\pm0.05$ mag as measured by \cite{Glass1975} at 3.5 $\mu$m, the companion would have a mass of $0.10-0.38\,M_\odot$ (epoch 1) and $0.19-0.63\,M_\odot$ (epoch 2) for ages in the $1-3$ Myr range and $L'$ extinctions in the aforementioned interval. These ranges include 1-$\sigma$ errorbars.\\
As calculated in Section \ref{section2}, R\,CrA appeared $\sim$0.22 mag brighter in the first observation than in the second one. Even if this offset were due to intrinsic stellar variation, the estimated contrasts (Table \ref{tab:astrometry}) indicate that the $L'$ brightness of the companion is subject to variability of $\Delta\text{mag}\approx0.4\pm0.23$ mag, making the mass estimation through model grids particularly difficult. Variability of this order of magnitude is expected for low mass young protostars \citep[e.g.,][]{Cody2014}, in particular if part of the flux is due to accretion processes occurring at varying rates \citep{Wolk2013} and emitting IR flux. Another process that might contribute to the observed $L'$ flux is the reprocessing of stellar  photons that are absorbed by circumstellar material and re-emitted in the mid-IR. Unfortunately, with the available data, it is not possible to estimate the fraction of reprocessed $L'$ flux. Accounting for the presence of accreting circumstellar material and of a dusty envelope, the estimated mass of the companion should be considered an upper limit.\\


\section{Summary and conclusions}
\label{section5}

As shown in several previous studies, the environment in the `Coronet' is extremely complex. Our NaCo high-contrast imaging data revealed an additional physical component of this region, which will need to be taken into account when interpreting the cloud environment harboring R\,CrA. Follow-up observations with a baseline of a few years will allow to better constrain the object's orbit and further investigate its time variability. High-contrast observations at shorter wavelengths may shed light on the object's spectral type and help quantify extinction effects. Finally, the presence of material accreting onto the companion candidate and emitting IR flux needs to be considered and investigated. In this case, different contributions to the IR emission should be disentangled and the object mass could be calculated from the true photospheric emission of the object.


\begin{acknowledgements}
The authors thank T. Stolker and C. Bailer-Jones for helpful and constructive discussions. Part of this work has been carried out within the frame of the National Center for Competence in Research PlanetS supported by the Swiss National Science Foundation. SPQ acknowledges the financial support of the SNSF. GC and SPQ thank the Swiss National Science Foundation for financial support under grant number 200021\_169131. GMK is supported by the Royal Society as a Royal Society University Research Fellow. AM acknowledges the support of the DFG priority program SPP 1992 "Exploring the Diversity of Extrasolar Planets"(MU 4172/1-1). JO and NG acknowledge financial support from the ICM (Iniciativa Cient\'ifica Milenio) via the N\'ucleo Milenio de Formaci\'on Planetaria grant. JO acknowledges support from the Universidad de Valpara\'iso, and from Fondecyt (grant 1180395). This research has made use of the SIMBAD database, operated at CDS, Strasbourg, France. This work has made use of data from the European Space Agency (ESA) mission {\it Gaia} (\url{https://www.cosmos.esa.int/gaia}), processed by the {\it Gaia} Data Processing and Analysis Consortium (DPAC, \url{https://www.cosmos.esa.int/web/gaia/dpac/consortium}). Funding for the DPAC has been provided by national institutions, in particular the institutions participating in the {\it Gaia} Multilateral Agreement. This publication makes use of data products from the Wide-field Infrared Survey Explorer, which is a joint project of the University of California, Los Angeles, and the Jet Propulsion Laboratory/California Institute of Technology, funded by the National Aeronautics and Space Administration.

\end{acknowledgements}

%
%

\bibliographystyle{aa}



\end{document}